\documentclass[twocolumn,superscriptaddress,preprintnumbers,10pt,aps,pra]{revtex4-1}
\usepackage{color}
\usepackage{graphicx}
\usepackage{amsmath,amssymb}
\usepackage{ragged2e}
\usepackage{keyval}
\usepackage[justification=centering]{subfig}
\usepackage[justification=RaggedRight,singlelinecheck=off]{caption}
\usepackage{braket}
\usepackage{natbib}
\usepackage[colorlinks,allcolors=blue]{hyperref}
\usepackage{breakurl}
\usepackage{textcase}
\usepackage{bm}
\newcommand{\ph}{\ensuremath{\varphi}}
\newcommand{\ii}{\ensuremath{\mathrm{i}}}
\newcommand{\ee}{\ensuremath{\mathrm{e}}}
\newcommand{\lt}{\ensuremath{\tilde{l}}}
\newcommand{\pt}{\ensuremath{\tilde{p}}}
\newcommand{\abs}[1]{\ensuremath{\left| #1 \right|}}
\newcommand{\norm}[1]{\ensuremath{\left\| #1 \right\|}}

\begin{document}
\title{Protecting OAM Photons from Decoherence in a Turbulent Atmosphere} 
\author{Jose Raul \surname{Gonzalez Alonso}}
\email[Electronic address: ]{jrgonzal@usc.edu}
\author{Todd A. Brun}
\email[Electronic address: ]{trbrun@usc.edu}
\affiliation{Department of Physics and Astronomy, University of Southern California, Los Angeles, California, 90089-0484, USA}
\begin{abstract}
One of the most important properties of orbital angular momentum (OAM) of photons is that the Hilbert space required to describe a general quantum state is infinite dimensional. In principle, this could allow
for encoding arbitrarily large amounts of quantum information per photon, but in practice, this potential is limited by decoherence and errors. To determine whether photons with OAM are suitable for quantum communication, we numerically simulated their passage through a turbulent atmosphere and the resulting errors. We also proposed an encoding scheme to protect the photons from these errors, and characterized its effectiveness by the channel fidelity.
\end{abstract}
\maketitle
\section{Introduction}
It is a well known fact that in classical electromagnetic theory the total angular momentum of a field can always be decomposed into the sum of two terms \cite{jackson_classical_1999}. One of these terms is identifiable as the
orbital part while the other as the polarization part of the angular momentum. Upon expressing these quantities as quantum mechanical operators \cite{mandel_wolf}, it becomes clear that each of them requires very different spaces
to be diagonalized. In the case of the polarization angular momentum, it is sufficient to have a two-dimensional Hilbert space. On the other hand, for the orbital case, an infinite dimensional space is needed.

Photons have always been the information carriers of choice in quantum information, with
many protocols using polarization or time-bin degrees of freedom to encode quantum
information \cite{PhysRevLett.81.3283, 919992}. Exploiting the photon's orbital angular momentum (OAM) could provide distinctive advantages. The main one 
is an increased alphabet size for information transmission \cite{PhysRevLett.88.013601}. This interesting property, could be applied in quantum key distribution \cite{Spedalieri2006340, PhysRevLett.88.127902}.
However, this potential can only be realized if suitable quantum information can be encoded in the OAM 
photon states, and if it can be protected from the decohering effect of atmospheric turbulence \cite{PhysRevA.45.8185, PhysRevLett.94.153901, Gbur:08, Tyler:09, Tyler:09, PhysRevA.74.013805, 1456150, 1451964}.

To better understand the decoherence a photon undergoes when traveling through the atmosphere we numerically simulated the evolution and extracted a Kraus operator sum 
decomposition for the resulting error process. We then studied one scheme for encoding
and correcting quantum information. 

The plan of this paper is as follows. The theoretical framework for the turbulence model,
and the calculation of the Kraus operators, are presented in sections \ref{sec:turb} and \ref{sec:num-proc}. In section \ref{sec:num-ex}, we
describe the results of our numerical simulations. In section \ref{sec:enc}, we discuss the problem of encoding information when we cannot do perfect error correction. In section \ref{sec:enc-res} we present the result of encoding
a qubit in a photon with OAM and how the channel fidelity behaves with the distance travelled for a given encoding. Finally, in section \ref{sec:concl} we conclude.

\section{OAM States and Atmospheric Turbulence}\label{sec:turb}
\begin{figure*}[!htbp]
\centering
\includegraphics[scale=1.0]{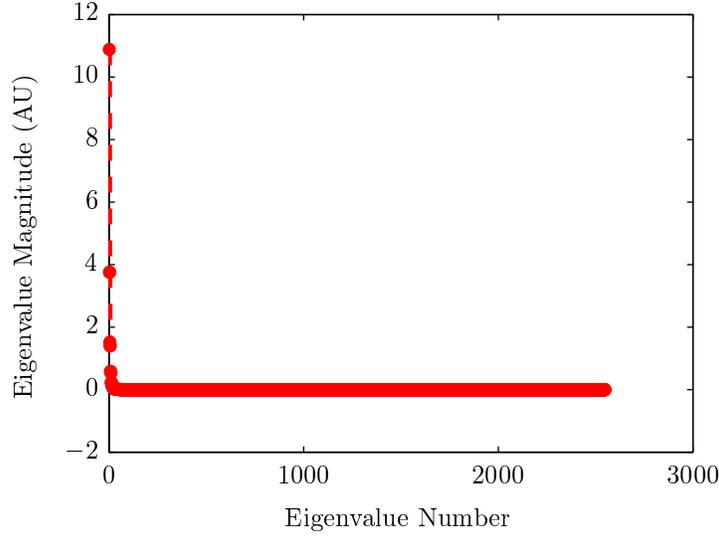}
\caption{(Color online) Example of the spectrum of $\mathbf{R}$ when $max_{in}=3$, $max_{out}=6$, $w_0=0.01$ m, $C_n^2=1\times 10^{-14}$ $\mbox{m}^{-2/3}$, $\lambda=1\times 10^{-6}$ m, $z=500$ m.}
\label{fig:eigsR}
\end{figure*}
Consider a beam that initially has a spatial wave function corresponding to an eigenstate of OAM. Write
such an eigenfunction in cylindrical coordinates as
    \begin{equation}\label{eq:wavefnOAM}
        \Braket{\mathbf{r}|l_0,p_0} = \frac{1}{\sqrt{2 \pi}}R_{l_0,p_0}(r,z)\exp(\ii l_0\theta),
    \end{equation}
where $r^2 = x^2 + y^2$, $\theta = \arctan\left(\frac{y}{x}\right)$, and the beam proppagates in the $z$ direction. The functions $R_{l_0,p_0}(r,z)$ are a basis for the radial dependence, such as the 
Laguerre-Gauss functions \cite{PhysRevA.45.8185}. Throughout this paper we use
\begin{eqnarray}
       R_{l_0,p_0}(r,z) &= \frac{A}{w(z)}\left(\frac{\sqrt{2} r}{w(z)}\right)^{\abs{l_0}} L_{p_0}^{\abs{l_0}}\left(\frac{2r^2}{w(z)^2}\right)\ee^{-r^2/w(z)^2}\nonumber\\
       &\times \ee^{-\ii kr^2/[2R(z)]}\ee^{ \ii(2p_0+\abs{l_0} + 1)\tan^{-1}(z/z_R)}, \label{eq:radialpartOAM}
\end{eqnarray}
where $w(z)=w_0\sqrt{1+(z/z_R)^2}$ is the beam width, $R(z)=z(1+(z_R/z)^2)$ is the radius of wavefront curvature, and $z_R = \frac{1}{2}kw_0^2$ is the
Rayleigh range. The quantity $\tan^{-1}(z/z_R)$ is known as the Gouy phase.

We assume that the effects of the atmospheric turbulence can be represented by the action of an operator 
$\hat{\mathrm{T}}_\ph$  such that
    \begin{equation}\label{eqn:turbeffect}
        \Big<\mathbf{r}\Big|\hat{\mathrm{T}}_{\ph}\Big|l_0,p_0\Big> = \exp\left(\ii \ph (r,\theta)\right)\Braket{\mathbf{r}|l_0,p_0}.
    \end{equation}
 That is, the spatially-varying phase change $\ph(r,\theta)$ represents the cumulative effect of fluctuations in the refractive index of air
 \cite{PhysRevLett.94.153901, 1367-2630-9-4-094}. In general, $\ph(r,z)$ will be a random function drawn from a suitable ensemble. The state of the beam
  after the interaction with the environment is a superposition of several OAM eigenstates:
    \begin{equation}\label{eqn:OAMsuperposition}
        \Ket{l_0,p_0} \overset{\hat{\mathrm{T}}_{\ph}}{\mapsto} \sum_{l=0}^{\infty}\sum_{p=0}^{\infty} \alpha_{l,p,l_0,p_0}\Ket{l,p}.
    \end{equation}
The coefficients $\left\{ \alpha_{l,p,l_0,p_0} \right\}$ are given by
    \begin{eqnarray}\label{eqn:OAMexpcoef}
        &&\alpha_{l,p,l_0,p_0} = \frac{1}{2\pi}\iint r \;\mathrm{d}r\,\mathrm{d}\theta\;
        \overline{R_{l,p}}(r,z)\nonumber\\
        &\times&\exp(\ii\left( \theta \left( l_0-l \right) - \ph(r,\theta) \right)) R_{l_0,p_0}(r,z).
    \end{eqnarray}

We are interested in studying the effect of turbulence on the most general possible quantum state. By linearity, it is sufficient to consider initial operators of the 
form $\Ket{l,p}\Bra{l',p'}$. These operators transform according to
    \begin{equation}\label{eqn:generalturbeffect}
        \Ket{l,p}\Bra{l',p'} \mapsto \hat{\mathrm{T}}_{\ph}\Ket{l,p}\Bra{l',p'}\hat{\mathrm{T}}_{\ph}^\dagger.
    \end{equation}
From equations (\ref{eqn:OAMsuperposition}--\ref{eqn:generalturbeffect}) it follows that
    \begin{eqnarray}
        &&\Ket{l,p}\Bra{l',p'} \mapsto \frac{1}{4\pi^2} \sum_{\lt,\pt,\lt',\pt'} \iiiint r \;\mathrm{d}r\,\mathrm{d}\theta\; r'\;\mathrm{d}r'\,\mathrm{d}\theta'\nonumber\\
        &\times& \overline{R_{\lt,\pt}}(r,z) R_{l,p}(r,z) \exp\left(\ii\left( \theta \left( l-\lt \right) + \ph(r,\theta) \right)\right) \nonumber\\
        &\times& R_{\lt',\pt'}(r',z) \overline{R_{l',p}}(r',z) \exp\left(-\ii\left( \theta' \left( l'-\lt' \right) + \ph(r',\theta') \right)\right)\nonumber\\
        &\times& \Ket{\lt,\pt}\Bra{\lt',\pt'}.\label{eqn:genturbeffect}
    \end{eqnarray}
Since the variations in the atmosphere are random, we take their ensemble average.
If one assumes that the refractive index fluctuations in the atmosphere are a Gaussian random process with zero mean, then one can use that 
$\Braket{\exp(\ii x)} = \exp\left( - \frac{1}{2} \Braket{x^2} \right)$ and Eq.
\eqref{eqn:genturbeffect} to show that on average a general OAM state undergoes the following transformation
    \begin{eqnarray}
        &&\Ket{l,p}\Bra{l',p'} \mapsto \frac{1}{2\pi} \sum_{\lt,\pt,\pt'} \iiint r \;\mathrm{d}r\,\mathrm{d}\mu\;r'\;\mathrm{d}r'\;
        \overline{R_{\lt,\pt}}(r,z)\nonumber\\
        &\times& R_{l,p}(r,z) R_{l + \lt + l',\pt'}(r',z) \overline{R_{l',p'}}(r',z)\exp\left(\ii\mu\left(l-\lt\right)\right)\nonumber\\
        &\times& \exp\left( -\frac{\mathcal{D}_\ph \left(\left| \mathbf{r}-\mathbf{r'} \right|\right)}{2} \right) \Ket{\lt,\pt}\Bra{l' + \lt - l,\pt'}.\label{eq:meanturbeffect}
    \end{eqnarray}
In the equation above $\mathcal{D}_\ph$ is called the phase structure function of the aberrations. We assume that this function is rotationally invariant, 
that is, the aberrations are isotropic, so that 
$$\mathcal{D}_\ph \left(\left| \mathbf{r}-\mathbf{r'} \right|\right) = \mathcal{D}_\ph (r,\theta,r',\theta') = \mathcal{D}_\ph (r,\theta-\theta',r',0).$$ 
This, along with the trivial change of variables $\mu = \theta-\theta', \nu = \frac{1}{2} (\theta+\theta')$ turns Eq. \eqref{eqn:genturbeffect} into Eq.
\eqref{eq:meanturbeffect}. This phase structure function depends on the model of turbulence and the power spectral density of the fluctuations 
\cite{Andrews:LaserProp, 2010arXiv1009.1956R}. For the Kolmogorov turbulence theory, the function $\mathcal{D}_\ph$ is given by
 \cite{FRIED:65, FRIED:66}
    \begin{equation}
        \mathcal{D}_\ph \left(\left| \mathbf{r}-\mathbf{r'} \right|\right) = 2\left(\frac{24}{5}\Gamma\left( \frac{6}{5}\right)\right)^{\frac{5}{6}} \left(\frac{\left|\mathbf{r}-\mathbf{r'} \right|}{r_0}\right)^{\frac{5}{3}},
    \end{equation}
where
\begin{equation}
r_0 = \left(\frac{16.6}{\lambda^2}\int_{L} \mathrm{d}\ell\, C_n^2\right)^{-\frac{3}{5}}
\label{eq:friedparameter}
\end{equation}
is the Fried parameter, which has dimensions of length \cite{FRIED:65}.  In \eqref{eq:friedparameter},
$\lambda$ is the wavelength, $L$ is the propagation path, $\ell$ is length along the propagation path, and $C_n$ is called the atmospheric refractive index 
structure constant. In spite of being called a constant, $C_n$ depends on altitude and may vary along the path \cite{FRIED:65}. Note that other models for 
turbulence may lead to different expressions \cite{Andrews:LaserProp}.

By evaluating \eqref{eq:meanturbeffect}, we can extract the superoperator  elements of a completely positive, trace-preserving map $\hat{\mathcal{T}}_\ph$.  
This map represents the decoherence process of the OAM states. 
By analyzing $\hat{\mathcal{T}}_\ph$, we hope to find the dominant noise processes of this decoherence, and construct error-correction procedures
that protect against them.  Both the integral and the analysis of the superoperator must in general be done numerically.

\section{Numerical Simulation: Procedure}\label{sec:num-proc}
\begin{figure*}[!htbp]
\centering
\subfloat{\includegraphics[scale=1.0]{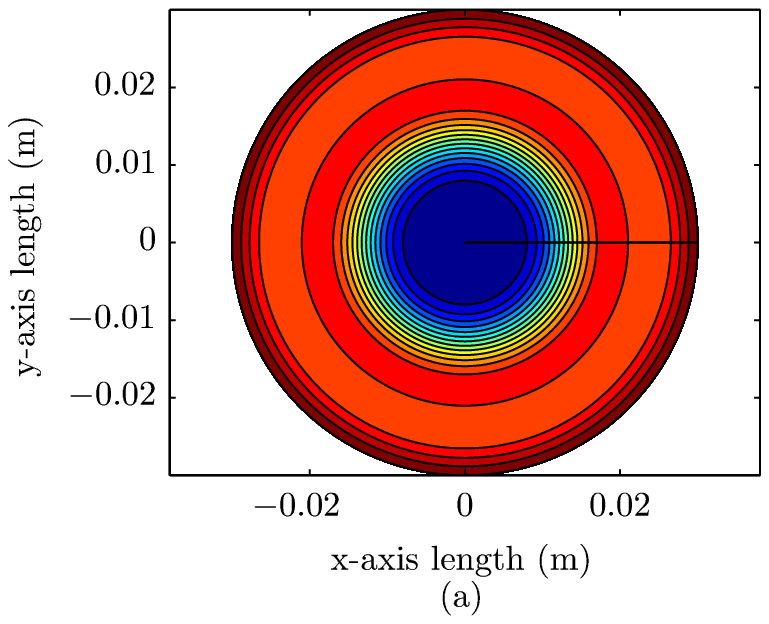}}
\hfill
\subfloat{\includegraphics[scale=1.0]{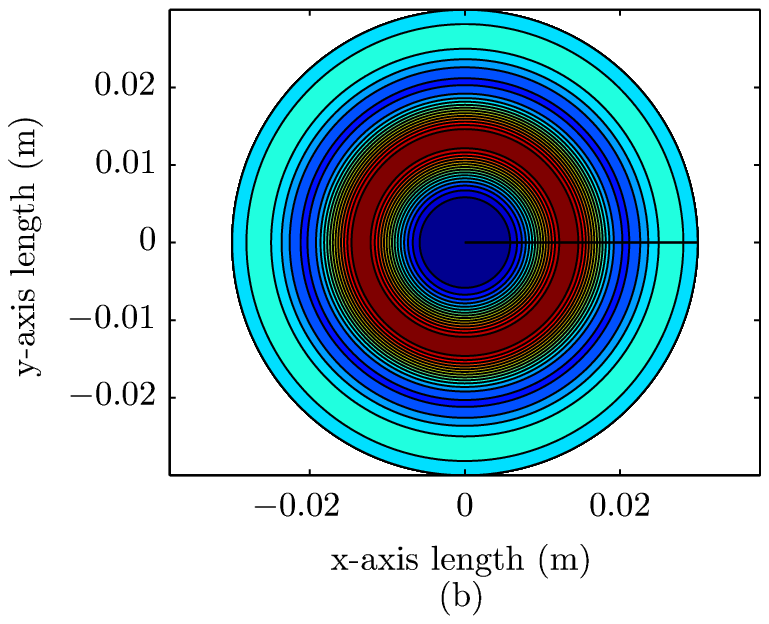}}
\caption{ (Color online) Contour plots of (a) $\ket{0_L}$ and (b) $\ket{1_L}$ before the noise process, where $max_{in}=3$, $max_{out}=6$, $w_0=0.01$ m, $C_n^2=1\times 10^{-14}$ $\mbox{m}^{-2/3}$, $\lambda=1\times 10^{-6}$ m, and $z=500$ m.}
\label{fig:log-without-noise}
\end{figure*}
\begin{figure*}[!htbp]
\centering
\subfloat{\includegraphics[scale=1.0]{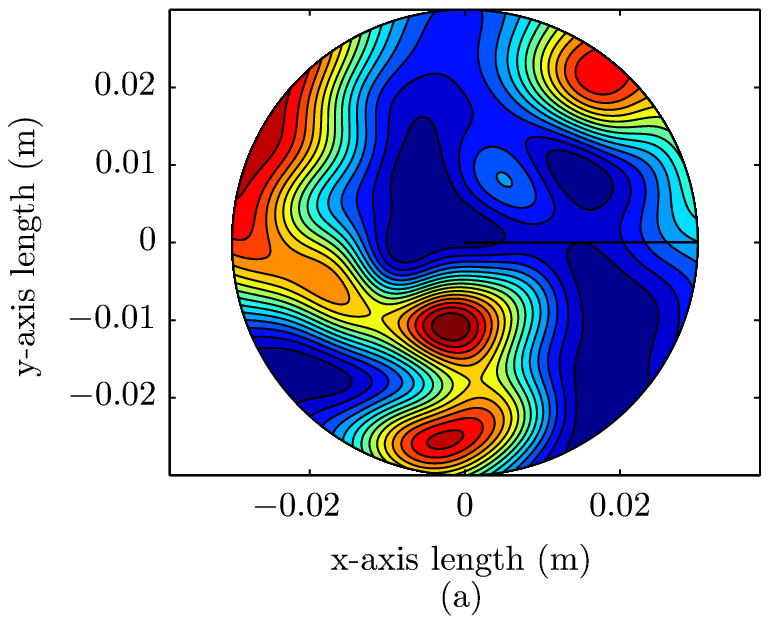}}
\hfill
\subfloat{\includegraphics[scale=1.0]{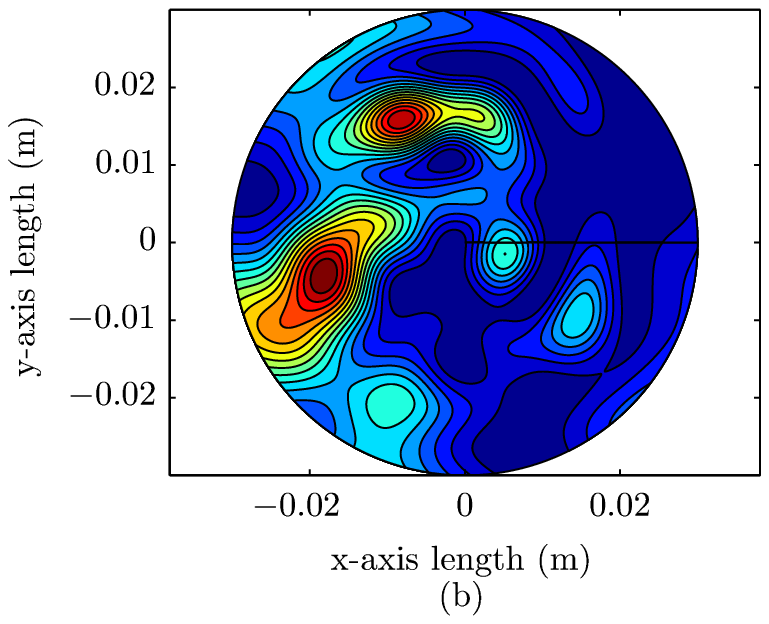}}
\caption{ (Color online) Contour plots of (a) $\ket{0_L}$ and (b) $\ket{1_L}$ after the noise process, where $max_{in}=3$, $max_{out}=6$, $w_0=0.01$ m, $C_n^2=1\times 10^{-14}$ $\mbox{m}^{-2/3}$, $\lambda=1\times 10^{-6}$ m, and $z=500$ m.}
\label{fig:log-with-noise}
\end{figure*}
\begin{figure*}[!htbp]
\centering
\subfloat{\includegraphics[scale=1.0]{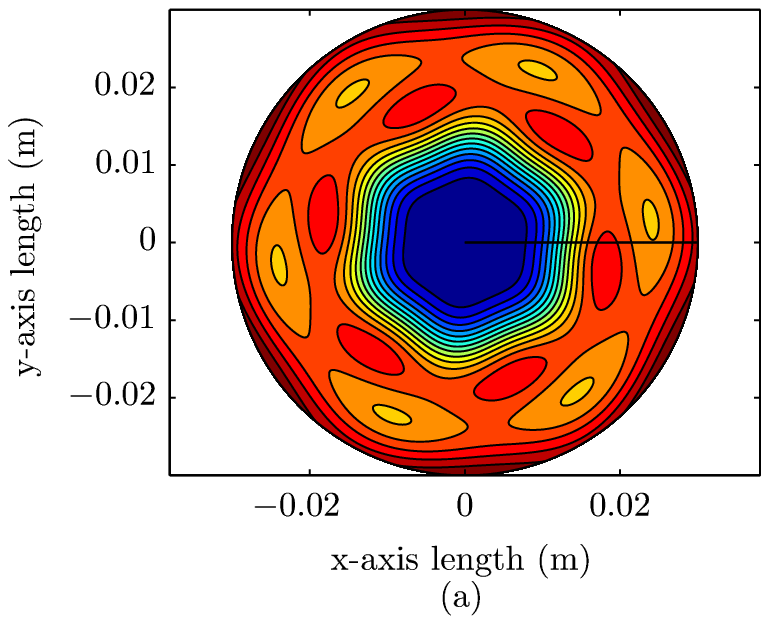}}
\hfill
\subfloat{\includegraphics[scale=1.0]{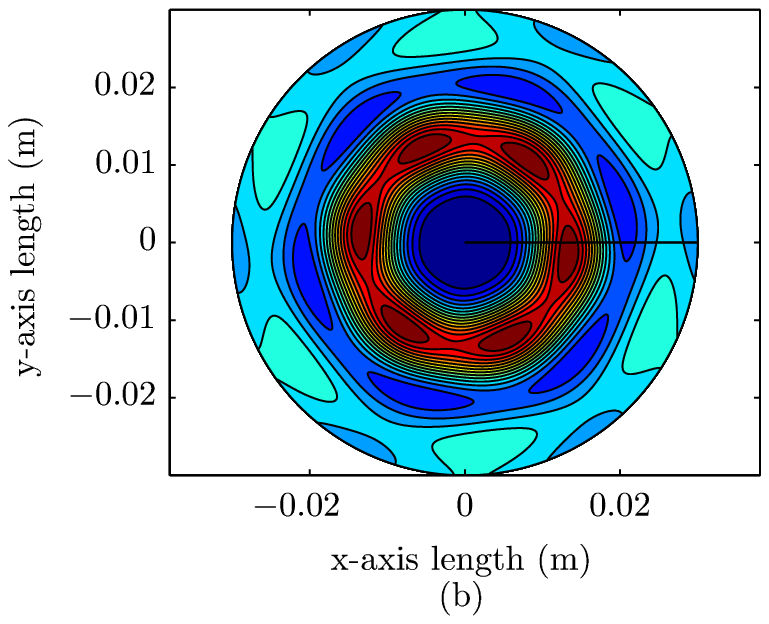}}
\caption{ (Color online) Contour plots of (a) $\ket{0_L}$ and (b) $\ket{1_L}$ after the recovery process, where $max_{in}=3$, $max_{out}=6$, $w_0=0.01$ m, $C_n^2=1\times 10^{-14}$ $\mbox{m}^{-2/3}$, $\lambda=1\times 10^{-6}$ m, and $z=500$ m.}
\label{fig:log-rec}
\end{figure*}
The elements of the superoperator $\hat{\mathcal{T}}_\ph$ can be represented as a matrix. 
Taking a cue from expression \eqref{eqn:genturbeffect}, the matrix elements of $\mathbf{T}$ should satisfy 
 \begin{equation}\label{eqn:matrix_rep_T}
 \left(\hat{\mathcal{T}}_\ph \rho \right)_{(\lt,\pt,\lt',\pt')} = \sum_{l,p,l',p'} \mathbf{T}_{(\lt,\pt,\lt',\pt'),(l,p,l',p')} \rho_{(l,p,l',p')},
 \end{equation}
 where we represent density operators as vectors in an appropriately chosen Hilbert space.  The elements of the vector are
 \begin{equation}\label{eqn:rho_rep}
 \rho_{(l,p,l',p')} =  \Braket{ l,p | \rho | l',p'}
 \end{equation}
 With this choice, the matrix elements of $\mathbf{T}$ are
 \begin{equation}\label{eqn:matrixT}
 \mathbf{T}_{(\lt,\pt,\lt',\pt'),(l,p,l',p')}  = \begin{cases} 0, & \mbox{if } \lt'\ne l' + \lt - l \\ \iiint \mbox{in \eqref{eq:meanturbeffect}}, & \mbox{if } \lt' = l' + \lt - l 
 \end{cases}.
 \end{equation}
 
As mentioned in the previous section, the matrix elements of $\mathbf{T}$ in general must be evaluated numerically . For this 
purpose,  we have used the integration algorithms in \cite{Genz1980295} and \cite{Berntsen:1991:AAA:210232.210233} as implemented by Steven G. 
Johson \footnote{\url{http://ab-initio.mit.edu/wiki/index.php/Cubature}}, and the GNU Scientific Library \cite{GSL}.

In principle, the states used for communication could be arbitrary superpositions of 
OAM states. However, we will assume that the input states
to the channel are restricted to a subspace of the OAM Hilbert space with angular
momentum quantum numbers below some maximum. This kind of truncation of the Hilbert space
also makes the analysis easier; but since the effects of turbulence can change the OAM of
a photon, it creates the possibility of leakage errors, where the input photons are
transformed out of the subspace. To minimize this problem, we will allow the output
subspace of the photons to be higher-dimensional than the input subspace. That means that
the matrix of the superoperator can be rectangular.

In general, the matrix representation of the superoperator $\hat{\mathcal{T}}_\ph$ is sparse, and when considering different 
dimensions for the input and output spaces, rectangular. For the OAM states before the
noise process, we include states with orbital angular
momentum  projections $l_{in}$ such that $\abs{l_{in}}\le max_{in}$. After the noise
process, we include OAM states such that $\abs{l_{out}}\le max_{out}$. For the radial
quantum number, we include states $\ket{l,p}$ such that $0\le p \le max_{in}$ or 
$max_{out}$ respectively.

From the noise process represented by $\hat{\mathcal{T}}_\ph$, we can obtain a 
Kraus representation. To this end, consider the matrix $\mathbf{R}$ defined by
rearranging the elements of $\mathbf{T}$:
\begin{equation}\label{eqn:matrixR}
\mathbf{R}_{(\lt,\pt, l,p),(\lt',\pt',l',p')} =  \mathbf{T}_{(\lt,\pt,\lt',\pt'),(l,p,l',p')} .
\end{equation}
The matrix $\mathbf{R}$ is always square, regardless of the values of $max_{in}$ and
$max_{out}$ we use to truncate the input and output spaces. With this definition,
\begin{eqnarray*}
&&\rho \overset{\hat{\mathcal{T}}_{\ph}}{\mapsto}\sum_{\substack{l,p,l',p'\\ \lt,\pt,\lt',\pt'}} \mathbf{T}_{(\lt,\pt,\lt',\pt'),(l,p,l',p')} \rho_{(l,p,l',p')} \ket{\lt,\pt}\bra{\lt',\pt'}\nonumber\\
&=&\sum_{\substack{l,p,l',p'\\ \lt,\pt,\lt',\pt'}} \mathbf{R}_{(\lt,\pt, l,p),(\lt',\pt',l',p')} \hat{\mathrm{O}}_{(\lt,\pt)(l,p)} \rho \left(\hat{\mathrm{O}}_{(\lt',\pt')(l',p')}\right)^\dagger
\end{eqnarray*}
where $\hat{\mathrm{O}}_{(\lt,\pt)(l,p)} = \ket{\lt,\pt}\Bra{l,p}$. We now diagonalize $\mathbf{R}$:
\begin{equation}\label{eqn:diag_r}
\mathbf{R}_{(\lt,\pt, l,p),(\lt',\pt',l',p')} = \sum_k \lambda^k v_{(\lt,\pt,l,p)}^k \left(v_{(\lt',\pt',l',p')}^k\right)^*,
\end{equation}
and use this to obtain a Kraus decomposition
\begin{equation}\label{eqn:kraus_decomp}
\rho \overset{\hat{\mathcal{T}}_{\ph}}{\mapsto}\sum_k \hat{\mathrm{A}}_k \rho \hat{\mathrm{A}}_k^\dagger,
\end{equation} 
with Kraus operators $\hat{\mathrm{A}}_k$ given by
\begin{equation}\label{eqn:def_kraus_op}
\hat{\mathrm{A}}_k = \sum_{\lt,\pt,l,p} \sqrt{\lambda^k} v_{(\lt,\pt,l,p)}^k \hat{\mathrm{O}}_{(\lt,\pt)(l,p)}.
\end{equation}
The decomposition in \eqref{eqn:kraus_decomp} will allow us to order the Kraus operators by their respective importance, that 
is, by the size of their eigenvalues, and will also give us some intuition about the
dominant effects of the noise process on our initial state.

\section{Numerical Simulations: Examples}\label{sec:num-ex}
To illustrate our discussion so far, we present some of the results of the
numerical simulations needed to obtain the elements of $\mathbf{T}$, 
and $\mathbf{R}$ (see Eqs. \eqref{eqn:matrixT} and \eqref{eqn:matrixR}), and the
error operators defined by Eq. \eqref{eqn:def_kraus_op}.

First, we observe that the eigenspectrum of $\mathbf{R}$ is dominated by its 
few largest eigenvalues (see Fig. \ref{fig:eigsR}).
The first, and largest one, is for $\hat{\mathrm{A}}_1$.
Upon further inspection, we see that $\hat{\mathrm{A}}_1$ is close to the identity operator
\begin{equation}\label{eqn:error_op_A1}
\hat{\mathrm{A}}_1 = c\hat{\mathrm{I}} + \hat{\mathrm{\epsilon}}, 
\end{equation}
where $\norm{\hat{\mathrm{\epsilon}}}$ is small. We note that some of the remaining
nonzero eigenvalues are degenerate, corresponding to a pair of error operators.

The actions of two error operators with the same eigenvalue parallel each other. One
operator raises the initial value of $l$ by 
a given amount, while the other lowers it by the same amount. As the eigenvalues become
smaller and smaller, the values by which the pair of error operators raise and lower $l$
become larger and larger. Put differently, large changes in the value of OAM for a beam
are unlikely. The question of how to encode our initial OAM states to protect them from
the noise process will be studied in the following sections.

\section{Encoding and Protecting a Qubit}\label{sec:enc}
In this section we explore the possibility of encoding a qubit using a 
suitable basis of OAM states to protect it agaainst turbulence. The basis states should
be such that after encoding a qubit and then performing a suitable noise recovery
operation we maximize the quantum channel fidelity \cite{2011arXiv1106.1445W}
of the atmospheric turbulence channel. However, before choosing the encoding, we must
know how to implement the error recovery map.

Since we are truncating both the input and  output space we cannot expect to be able to
do perfect quantum error correction: the error map is not trace preserving. 
However, we can do \emph{approximate} quantum error correction (see
\cite{2009arXiv0909.0931K,2001quant.ph.12106S,1997PhRvA..56.2567L}). 
The idea behind approximate quantum error correction (AQEC for short) over a system with
Hilbert space $\mathcal{H}$ and noise process $\mathcal{E}$ is to solve the 
triple optimization problem
\begin{equation}\label{eqn:prob_aqec}
\max_{\mathcal{W}} \max_{\mathcal{R}} \min_{\ket{\psi}\in\mathcal{H}_0} F
\left(  \ket{\psi}\bra{\psi}, \mathcal{W}^{-1} \circ \mathcal{R} \circ \mathcal{E} \circ
\mathcal{W} \ket{\psi}\bra{\psi}\right).
\end{equation}
Here, $\mathcal{H}_0$ denotes the qudit space (a subspace of $\mathcal{H}$), $\mathcal{E}$ the error map, $\mathcal{W}$ the encoding map,
$\mathcal{R}$ the recovery map, and $F$ the fidelity. One can fix an encoding map an then try to find the solutions to the problem
\begin{equation}\label{eqn:simpler_aqec_prob}
\max_{\mathcal{R}} \min_{\ket{\psi}\in\mathcal{C}} F \left(  \ket{\psi}\bra{\psi},\mathcal{R} \circ \mathcal{E} \ket{\psi}\bra{\psi}\right),
\end{equation}
where $\mathcal{C}$ is the code space. Although \eqref{eqn:simpler_aqec_prob} can
be solved by convex optimization methods \cite{2008arXiv0810.2524T}, we can more easily use the
transpose channel to construct a near optimal recovery map \cite{2009arXiv0909.0931K}.
The recovery map we use is given by 
\begin{equation}\label{eqn:recovery_map}
\mathcal{R}_P=\sum_k P \hat{\mathrm{A}}_k^\dagger\mathcal{E}( P )^{-1/2}(\cdot)\mathcal{E}( P )^{-1/2}\hat{\mathrm{A}}_k P.
\end{equation}
In Eq. \eqref{eqn:recovery_map}, $P$ is the projector onto the code space, and
$\mathcal{E}(\cdot)=\sum_k \hat{\mathrm{A}}_k (\cdot) \hat{\mathrm{A}}_k^\dagger$
is the error map. Finally, the inverse of $\mathcal{E}( P )$ (actually, a pseudoinverse)
is taken over its support. 

Now that we know how to implement the error recovery map for a particular choice of code,
we can write an expression for the channel fidelity $\mathcal{C}$:
\begin{equation}\label{eq:opt-for-p}
	\mathcal{C} = \max_P \frac{1}{d^2}\sum_{k,l} \mathrm{tr}\left\| P \hat{\mathrm{A}}_k^\dagger\mathcal{E}( P )^{-1/2} \hat{\mathrm{A}}_l \right\|^2,
\end{equation}
where $P$ is the projector onto the code space, and $d$ is the dimension of the space we
are encoding. For a qubit, $d=2$.

\section{Encoding and Recovery Results}\label{sec:enc-res}
To illustrate the performance of the AQEC approach, we now show some examples of how the
channel fidelity varies with the path length of the path for a qubit ($d=2$) encoded using
OAM. Ideally, we should choose our code space to maximize the channel fidelity. However,
even with the simplifications of the AQEC, optimizing \eqref{eq:opt-for-p} has only been
feasible for very small examples. In these examples, the optimal encoding was usually one
that maximized the distance between the OAM values of the basis states. For this reason,
we used the following encoding to investigate how robust OAM states with AQEC are to
atmospheric noise:
\begin{align}
\ket{0}_L &= \frac{\ket{-max_{in},0} + \ket{-max_{in},1}}{\sqrt{2}} \label{eq:enc-0},\\
\ket{1}_L &= \frac{\ket{max_{in},max_{in}-1}+\ket{max_{in},max_{in}}}{\sqrt{2}}\label{eq:enc-one}.
\end{align}

In figures \ref{fig:log-without-noise} - \ref{fig:log-rec} we plot the basis states we
have chosen for certain cases before and after the noise process, and after the recovery.
We can see that the recovery map visibly restores the states, though not perfectly, to
their initial conditions. 

\begin{figure*}[!htbp]
\centering
\includegraphics[scale=0.35]{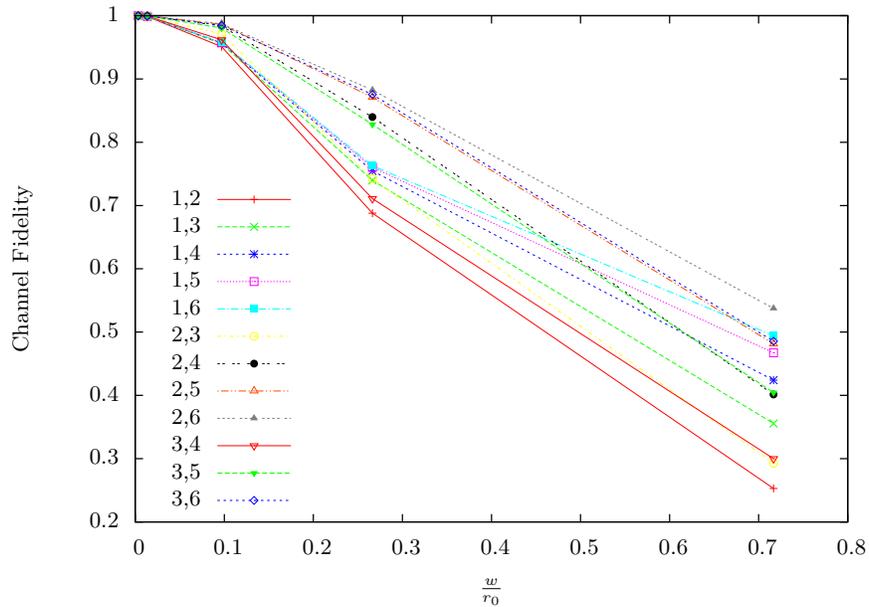}
\caption{(Color online) Behavior of the channel fidelity as a function of $w/r_0$ for several values of  $max_{in}$ and $max_{out}$. In all cases, $w_0=0.01$ m, $C_n^2=1\times 10^{-14}$ $\mbox{m}^{-2/3}$, $\lambda=1\times 10^{-6}$ m.}
\label{fig:cf-ratio-plot}
\end{figure*}

In our simulations we also investigated the effect of the dimensions of the input and
output spaces on the channel fidelity. For this purpose, we used $max_{in}=1,2,3$ for the
input space, and $max_{out}=max_{in},max_{in} + 1, \ldots, 6$ for the output space.
The output space was always at least as big as the input space. The effects of both
the dimensions and the path length are summarized in Fig. \ref{fig:cf-ratio-plot} and in Table
\ref{tab:cf-table}.
\begin{table*}[htbp]
\centering
\begin{tabular}{| c | c | c | c | c | c | c |}
\hline
$max_{in}$ & $max_{out}$ & $\frac{w}{r_0} = 3.4046\times 10^{-3}$ & $\frac{w}{r_0} = 1.3561\times 10^{-2}$ & $\frac{w}{r_0} = 9.6954\times 10^{-2}$ & $\frac{w}{r_0} = 2.6639\times 10^{-1}$ & $\frac{w}{r_0} = 7.1673\times 10^{-1}$ \\
\hline
1 &	1 &	0.9997 & 0.9965 & 0.8864 & 0.4749 & 0.1169 \\
1 &	2 &	0.9999 & 0.9987 & 0.9511 & 0.6879 & 0.2528 \\
1 &	3 &	0.9999 & 0.9988 & 0.9558 & 0.7399 & 0.3556 \\
1 &	4 &	0.9999 & 0.9988 & 0.9569 & 0.7548 & 0.4240 \\
1 &	5 &	0.9999 & 0.9988 & 0.9574 & 0.7603 & 0.4672 \\
1 &	6 &	0.9999 & 0.9988 & 0.9577 & 0.7629 & 0.4942 \\
2 &	2 &	0.9995 & 0.9954 & 0.8595 & 0.4659 & 0.1522 \\
2 &	3 &	0.9999 & 0.9994 & 0.9709 & 0.7423 & 0.2927 \\
2 &	4 &	1.0000 & 0.9997 & 0.9836 & 0.8399 & 0.4015 \\
2 &	5 &	1.0000 & 0.9997 & 0.9862 & 0.8710 & 0.4815 \\
2 &	6 &	1.0000 & 0.9998 & 0.9871 & 0.8824 & 0.5375 \\
3 &	3 &	0.9994 & 0.9938 & 0.8258 & 0.4344 & 0.1700 \\
3 &	4 &	0.9999 & 0.9992 & 0.9619 & 0.7110 & 0.3001 \\
3 &	5 &	1.0000 & 0.9995 & 0.9805 & 0.8282 & 0.4044 \\
3 &	6 &	1.0000 & 0.9996 & 0.9847 & 0.8754 & 0.4856 \\
\hline
\end{tabular}
\caption{The effect of the size of the input and output spaces on the channel fidelity when $w_0=0.01$ m, $C_n^2=1\times 10^{-14}$ $\mbox{m}^{-2/3}$, and $\lambda=1\times 10^{-6}$ m.}
\label{tab:cf-table}
\end{table*}


We expected that as we made the input space larger (higher values of $max_{in}$) the
values for the channel fidelity would improve, as there would be more distance between
the values of OAM in the encoded states. We see that in general terms this is true,
provided that the output space is also large enough to reduce leakage errors due to
truncation of the space (this is, for example, the reason why the case $max_{in}=2$,
$max_{out}=6$ has better channel fidelities than the case $max_{in}=3$, $max_{out}=6$).

Rather unsurprisingly, but disapointingly, we also see that the channel fidelity rapidly
decays as the ratio $w / r_0$ increases. This decay is so fast that, even with error correction,
after $w/r_0 = 2.6639\times 10^{-1}$, the OAM states for a moderate level of turbulence $C_n^2=1\times 10^{-14}\mbox{m}^{-2/3}$ 
become unusable for quantum communication. This could be an artifact of our
approximations, or of a suboptimal encoding or recovery. However, we believe it is more
likely that OAM photons are too sensitive to atmospheric turbulence to be useful for
quantum communication without some much better method of compensating for the noise.
Perhaps coupling quantum error correction with other protection methods, such as
decoherence-free subspaces, or the use of adaptive optics, may prove fruitful in
increasing the communications range over which OAM photons can be used to transmit
quantum information. These are subjects for future research.

Our results agree with the work done \cite{PhysRevA.74.062104, Tyler:09, 2010arXiv1009.1956R}
and, from an experimental perspective, from the behavior observed in \cite{2012arXiv1210.2867H,Malik:12,2013arXiv1301.7454R}. 
However, unlike these works, we have derived a Kraus representation for the atmospheric turbulence channel and studied the behavior
of the channel fidelity, not the channel capacity as is more commonly done. Also, we have studied the application of approximate quantum error
correction to the problem of protecting OAM photons from atmospheric turbulence, something that, as far as we know, has not been tried before.

\section{Conclusions}\label{sec:concl}
In principle, OAM could be used to encode large amounts of quantum information per photon
to be transmitted through air. Through numerical simulations, we were able to extract a
Kraus representation for the error process OAM photons undergo
through a turbulent atmosphere. We described the effects of turbulence using Kolmogorov's
model. To protect OAM photons form errors, we applied the methods of approximate quantum
error correction. Unfortunately, our numerical simulations indicate that even with
quantum error correction, the range over which we can use OAM for effective quantum
communications is very limited. Perhaps it is possible to better protect OAM photons
from noise if we couple quantum error correction with other methods like adaptive optics
or decoherence-free subspaces. These are the subject of ongoing research.

\begin{acknowledgments}
TAB and JRGA would like to thank Yan Yan, Yongxiong Ren, Nisar Ahmed, Yang Yue, Daniel 
Lidar, Paolo Zanardi, and Alan Willner for useful conversations. This work was supported
in part by NSF grants EMT-0829870 and TF-0830801. Computation for the work described in this paper
was supported by the University of Southern California Center for High-Performance Computing and 
Communications (\url{hpcc.usc.edu}).
\end{acknowledgments}

\bibliographystyle{apsrev4-1}
\bibliography{draft_refs}

\begin{thebibliography}{32}%
\makeatletter
\providecommand \@ifxundefined [1]{%
 \@ifx{#1\undefined}
}%
\providecommand \@ifnum [1]{%
 \ifnum #1\expandafter \@firstoftwo
 \else \expandafter \@secondoftwo
 \fi
}%
\providecommand \@ifx [1]{%
 \ifx #1\expandafter \@firstoftwo
 \else \expandafter \@secondoftwo
 \fi
}%
\providecommand \natexlab [1]{#1}%
\providecommand \enquote  [1]{``#1''}%
\providecommand \bibnamefont  [1]{#1}%
\providecommand \bibfnamefont [1]{#1}%
\providecommand \citenamefont [1]{#1}%
\providecommand \href@noop [0]{\@secondoftwo}%
\providecommand \href [0]{\begingroup \@sanitize@url \@href}%
\providecommand \@href[1]{\@@startlink{#1}\@@href}%
\providecommand \@@href[1]{\endgroup#1\@@endlink}%
\providecommand \@sanitize@url [0]{\catcode `\\12\catcode `\$12\catcode
  `\&12\catcode `\#12\catcode `\^12\catcode `\_12\catcode `\%12\relax}%
\providecommand \@@startlink[1]{}%
\providecommand \@@endlink[0]{}%
\providecommand \url  [0]{\begingroup\@sanitize@url \@url }%
\providecommand \@url [1]{\endgroup\@href {#1}{\urlprefix }}%
\providecommand \urlprefix  [0]{URL }%
\providecommand \Eprint [0]{\href }%
\providecommand \doibase [0]{http://dx.doi.org/}%
\providecommand \selectlanguage [0]{\@gobble}%
\providecommand \bibinfo  [0]{\@secondoftwo}%
\providecommand \bibfield  [0]{\@secondoftwo}%
\providecommand \translation [1]{[#1]}%
\providecommand \BibitemOpen [0]{}%
\providecommand \bibitemStop [0]{}%
\providecommand \bibitemNoStop [0]{.\EOS\space}%
\providecommand \EOS [0]{\spacefactor3000\relax}%
\providecommand \BibitemShut  [1]{\csname bibitem#1\endcsname}%
\let\auto@bib@innerbib\@empty
\bibitem [{\citenamefont {Jackson}(1999)}]{jackson_classical_1999}%
  \BibitemOpen
  \bibfield  {author} {\bibinfo {author} {\bibfnamefont {J.~D.}\ \bibnamefont
  {Jackson}},\ }\href {http://www.worldcat.org/oclc/774042102} {\emph {\bibinfo
  {title} {Classical electrodynamics}}},\ \bibinfo {edition} {3rd}\ ed.\
  (\bibinfo  {publisher} {Wiley},\ \bibinfo {address} {New York, {NY}},\
  \bibinfo {year} {1999})\BibitemShut {NoStop}%
\bibitem [{\citenamefont {Mandel}\ and\ \citenamefont
  {Wolf}(1995)}]{mandel_wolf}%
  \BibitemOpen
  \bibfield  {author} {\bibinfo {author} {\bibfnamefont {L.}~\bibnamefont
  {Mandel}}\ and\ \bibinfo {author} {\bibfnamefont {E.}~\bibnamefont {Wolf}},\
  }\href {http://www.worldcat.org/oclc/29564330} {\emph {\bibinfo {title}
  {{Optical Coherence and Quantum Optics}}}},\ \bibinfo {edition} {1st}\ ed.\
  (\bibinfo  {publisher} {Cambridge University Press},\ \bibinfo {year}
  {1995})\BibitemShut {NoStop}%
\bibitem [{\citenamefont {Buttler}\ \emph {et~al.}(1998)\citenamefont
  {Buttler}, \citenamefont {Hughes}, \citenamefont {Kwiat}, \citenamefont
  {Lamoreaux}, \citenamefont {Luther}, \citenamefont {Morgan}, \citenamefont
  {Nordholt}, \citenamefont {Peterson},\ and\ \citenamefont
  {Simmons}}]{PhysRevLett.81.3283}%
  \BibitemOpen
  \bibfield  {author} {\bibinfo {author} {\bibfnamefont {W.~T.}\ \bibnamefont
  {Buttler}}, \bibinfo {author} {\bibfnamefont {R.~J.}\ \bibnamefont {Hughes}},
  \bibinfo {author} {\bibfnamefont {P.~G.}\ \bibnamefont {Kwiat}}, \bibinfo
  {author} {\bibfnamefont {S.~K.}\ \bibnamefont {Lamoreaux}}, \bibinfo {author}
  {\bibfnamefont {G.~G.}\ \bibnamefont {Luther}}, \bibinfo {author}
  {\bibfnamefont {G.~L.}\ \bibnamefont {Morgan}}, \bibinfo {author}
  {\bibfnamefont {J.~E.}\ \bibnamefont {Nordholt}}, \bibinfo {author}
  {\bibfnamefont {C.~G.}\ \bibnamefont {Peterson}}, \ and\ \bibinfo {author}
  {\bibfnamefont {C.~M.}\ \bibnamefont {Simmons}},\ }\href {\doibase
  10.1103/PhysRevLett.81.3283} {\bibfield  {journal} {\bibinfo  {journal}
  {Phys. Rev. Lett.}\ }\textbf {\bibinfo {volume} {81}},\ \bibinfo {pages}
  {3283} (\bibinfo {year} {1998})}\BibitemShut {NoStop}%
\bibitem [{\citenamefont {Rarity}\ \emph {et~al.}(2001)\citenamefont {Rarity},
  \citenamefont {Gorman},\ and\ \citenamefont {Tapster}}]{919992}%
  \BibitemOpen
  \bibfield  {author} {\bibinfo {author} {\bibfnamefont {J.}~\bibnamefont
  {Rarity}}, \bibinfo {author} {\bibfnamefont {P.}~\bibnamefont {Gorman}}, \
  and\ \bibinfo {author} {\bibfnamefont {P.~R.}\ \bibnamefont {Tapster}},\
  }\href {\doibase 10.1049/el:20010334} {\bibfield  {journal} {\bibinfo
  {journal} {Electronics Letters}\ }\textbf {\bibinfo {volume} {37}},\ \bibinfo
  {pages} {512} (\bibinfo {year} {2001})}\BibitemShut {NoStop}%
\bibitem [{\citenamefont {Molina-Terriza}\ \emph {et~al.}(2001)\citenamefont
  {Molina-Terriza}, \citenamefont {Torres},\ and\ \citenamefont
  {Torner}}]{PhysRevLett.88.013601}%
  \BibitemOpen
  \bibfield  {author} {\bibinfo {author} {\bibfnamefont {G.}~\bibnamefont
  {Molina-Terriza}}, \bibinfo {author} {\bibfnamefont {J.~P.}\ \bibnamefont
  {Torres}}, \ and\ \bibinfo {author} {\bibfnamefont {L.}~\bibnamefont
  {Torner}},\ }\href {\doibase 10.1103/PhysRevLett.88.013601} {\bibfield
  {journal} {\bibinfo  {journal} {Phys. Rev. Lett.}\ }\textbf {\bibinfo
  {volume} {88}},\ \bibinfo {pages} {013601} (\bibinfo {year}
  {2001})}\BibitemShut {NoStop}%
\bibitem [{\citenamefont {Spedalieri}(2006)}]{Spedalieri2006340}%
  \BibitemOpen
  \bibfield  {author} {\bibinfo {author} {\bibfnamefont {F.~M.}\ \bibnamefont
  {Spedalieri}},\ }\href {\doibase 10.1016/j.optcom.2005.10.001} {\bibfield
  {journal} {\bibinfo  {journal} {Optics Communications}\ }\textbf {\bibinfo
  {volume} {260}},\ \bibinfo {pages} {340 } (\bibinfo {year}
  {2006})}\BibitemShut {NoStop}%
\bibitem [{\citenamefont {Cerf}\ \emph {et~al.}(2002)\citenamefont {Cerf},
  \citenamefont {Bourennane}, \citenamefont {Karlsson},\ and\ \citenamefont
  {Gisin}}]{PhysRevLett.88.127902}%
  \BibitemOpen
  \bibfield  {author} {\bibinfo {author} {\bibfnamefont {N.~J.}\ \bibnamefont
  {Cerf}}, \bibinfo {author} {\bibfnamefont {M.}~\bibnamefont {Bourennane}},
  \bibinfo {author} {\bibfnamefont {A.}~\bibnamefont {Karlsson}}, \ and\
  \bibinfo {author} {\bibfnamefont {N.}~\bibnamefont {Gisin}},\ }\href
  {\doibase 10.1103/PhysRevLett.88.127902} {\bibfield  {journal} {\bibinfo
  {journal} {Phys. Rev. Lett.}\ }\textbf {\bibinfo {volume} {88}},\ \bibinfo
  {pages} {127902} (\bibinfo {year} {2002})}\BibitemShut {NoStop}%
\bibitem [{\citenamefont {Allen}\ \emph {et~al.}(1992)\citenamefont {Allen},
  \citenamefont {Beijersbergen}, \citenamefont {Spreeuw},\ and\ \citenamefont
  {Woerdman}}]{PhysRevA.45.8185}%
  \BibitemOpen
  \bibfield  {author} {\bibinfo {author} {\bibfnamefont {L.}~\bibnamefont
  {Allen}}, \bibinfo {author} {\bibfnamefont {M.~W.}\ \bibnamefont
  {Beijersbergen}}, \bibinfo {author} {\bibfnamefont {R.~J.~C.}\ \bibnamefont
  {Spreeuw}}, \ and\ \bibinfo {author} {\bibfnamefont {J.~P.}\ \bibnamefont
  {Woerdman}},\ }\href {\doibase 10.1103/PhysRevA.45.8185} {\bibfield
  {journal} {\bibinfo  {journal} {Phys. Rev. A}\ }\textbf {\bibinfo {volume}
  {45}},\ \bibinfo {pages} {8185} (\bibinfo {year} {1992})}\BibitemShut
  {NoStop}%
\bibitem [{\citenamefont {Paterson}(2005)}]{PhysRevLett.94.153901}%
  \BibitemOpen
  \bibfield  {author} {\bibinfo {author} {\bibfnamefont {C.}~\bibnamefont
  {Paterson}},\ }\href {\doibase 10.1103/PhysRevLett.94.153901} {\bibfield
  {journal} {\bibinfo  {journal} {Phys. Rev. Lett.}\ }\textbf {\bibinfo
  {volume} {94}},\ \bibinfo {pages} {153901} (\bibinfo {year}
  {2005})}\BibitemShut {NoStop}%
\bibitem [{\citenamefont {Gbur}\ and\ \citenamefont {Tyson}(2008)}]{Gbur:08}%
  \BibitemOpen
  \bibfield  {author} {\bibinfo {author} {\bibfnamefont {G.}~\bibnamefont
  {Gbur}}\ and\ \bibinfo {author} {\bibfnamefont {R.~K.}\ \bibnamefont
  {Tyson}},\ }\href {http://josaa.osa.org/abstract.cfm?URI=josaa-25-1-225}
  {\bibfield  {journal} {\bibinfo  {journal} {J. Opt. Soc. Am. A}\ }\textbf
  {\bibinfo {volume} {25}},\ \bibinfo {pages} {225} (\bibinfo {year}
  {2008})}\BibitemShut {NoStop}%
\bibitem [{\citenamefont {Tyler}\ and\ \citenamefont {Boyd}(2009)}]{Tyler:09}%
  \BibitemOpen
  \bibfield  {author} {\bibinfo {author} {\bibfnamefont {G.~A.}\ \bibnamefont
  {Tyler}}\ and\ \bibinfo {author} {\bibfnamefont {R.~W.}\ \bibnamefont
  {Boyd}},\ }\href {http://ol.osa.org/abstract.cfm?URI=ol-34-2-142} {\bibfield
  {journal} {\bibinfo  {journal} {Opt. Lett.}\ }\textbf {\bibinfo {volume}
  {34}},\ \bibinfo {pages} {142} (\bibinfo {year} {2009})}\BibitemShut
  {NoStop}%
\bibitem [{\citenamefont {Berman}\ and\ \citenamefont
  {Chumak}(2006)}]{PhysRevA.74.013805}%
  \BibitemOpen
  \bibfield  {author} {\bibinfo {author} {\bibfnamefont {G.~P.}\ \bibnamefont
  {Berman}}\ and\ \bibinfo {author} {\bibfnamefont {A.~A.}\ \bibnamefont
  {Chumak}},\ }\href {\doibase 10.1103/PhysRevA.74.013805} {\bibfield
  {journal} {\bibinfo  {journal} {Phys. Rev. A}\ }\textbf {\bibinfo {volume}
  {74}},\ \bibinfo {pages} {013805} (\bibinfo {year} {2006})}\BibitemShut
  {NoStop}%
\bibitem [{\citenamefont {Fante}(1980)}]{1456150}%
  \BibitemOpen
  \bibfield  {author} {\bibinfo {author} {\bibfnamefont {R.}~\bibnamefont
  {Fante}},\ }\href {\doibase 10.1109/PROC.1980.11882} {\bibfield  {journal}
  {\bibinfo  {journal} {Proceedings of the IEEE}\ }\textbf {\bibinfo {volume}
  {68}},\ \bibinfo {pages} {1424} (\bibinfo {year} {1980})}\BibitemShut
  {NoStop}%
\bibitem [{\citenamefont {Fante}(1975)}]{1451964}%
  \BibitemOpen
  \bibfield  {author} {\bibinfo {author} {\bibfnamefont {R.}~\bibnamefont
  {Fante}},\ }\href {\doibase 10.1109/PROC.1975.10035} {\bibfield  {journal}
  {\bibinfo  {journal} {Proceedings of the IEEE}\ }\textbf {\bibinfo {volume}
  {63}},\ \bibinfo {pages} {1669} (\bibinfo {year} {1975})}\BibitemShut
  {NoStop}%
\bibitem [{\citenamefont {Gopaul}\ and\ \citenamefont
  {Andrews}(2007)}]{1367-2630-9-4-094}%
  \BibitemOpen
  \bibfield  {author} {\bibinfo {author} {\bibfnamefont {C.}~\bibnamefont
  {Gopaul}}\ and\ \bibinfo {author} {\bibfnamefont {R.}~\bibnamefont
  {Andrews}},\ }\href {http://stacks.iop.org/1367-2630/9/i=4/a=094} {\bibfield
  {journal} {\bibinfo  {journal} {New Journal of Physics}\ }\textbf {\bibinfo
  {volume} {9}},\ \bibinfo {pages} {94} (\bibinfo {year} {2007})}\BibitemShut
  {NoStop}%
\bibitem [{\citenamefont {Andrews}\ and\ \citenamefont
  {Phillips}(2005)}]{Andrews:LaserProp}%
  \BibitemOpen
  \bibfield  {author} {\bibinfo {author} {\bibfnamefont {L.~C.}\ \bibnamefont
  {Andrews}}\ and\ \bibinfo {author} {\bibfnamefont {R.~L.}\ \bibnamefont
  {Phillips}},\ }\href {http://www.worldcat.org/oclc/60767016} {\emph {\bibinfo
  {title} {Laser Beam Propagation through Random Media}}},\ \bibinfo {edition}
  {2nd}\ ed.\ (\bibinfo  {publisher} {SPIE Press},\ \bibinfo {year}
  {2005})\BibitemShut {NoStop}%
\bibitem [{\citenamefont {Roux}(2010)}]{2010arXiv1009.1956R}%
  \BibitemOpen
  \bibfield  {author} {\bibinfo {author} {\bibfnamefont {F.~S.}\ \bibnamefont
  {Roux}},\ }\href@noop {} {\bibfield  {journal} {\bibinfo  {journal} {ArXiv
  e-prints}\ } (\bibinfo {year} {2010})},\ \Eprint
  {http://arxiv.org/abs/1009.1956} {arXiv:1009.1956 [physics.optics]}
  \BibitemShut {NoStop}%
\bibitem [{\citenamefont {Fried}(1965)}]{FRIED:65}%
  \BibitemOpen
  \bibfield  {author} {\bibinfo {author} {\bibfnamefont {D.~L.}\ \bibnamefont
  {Fried}},\ }\href
  {http://www.opticsinfobase.org/abstract.cfm?URI=josa-55-11-1427} {\bibfield
  {journal} {\bibinfo  {journal} {J. Opt. Soc. Am.}\ }\textbf {\bibinfo
  {volume} {55}},\ \bibinfo {pages} {1427} (\bibinfo {year}
  {1965})}\BibitemShut {NoStop}%
\bibitem [{\citenamefont {Fried}(1966)}]{FRIED:66}%
  \BibitemOpen
  \bibfield  {author} {\bibinfo {author} {\bibfnamefont {D.~L.}\ \bibnamefont
  {Fried}},\ }\href
  {http://www.opticsinfobase.org/abstract.cfm?URI=josa-56-10-1372} {\bibfield
  {journal} {\bibinfo  {journal} {J. Opt. Soc. Am.}\ }\textbf {\bibinfo
  {volume} {56}},\ \bibinfo {pages} {1372} (\bibinfo {year}
  {1966})}\BibitemShut {NoStop}%
\bibitem [{\citenamefont {Genz}\ and\ \citenamefont
  {Malik}(1980)}]{Genz1980295}%
  \BibitemOpen
  \bibfield  {author} {\bibinfo {author} {\bibfnamefont {A.}~\bibnamefont
  {Genz}}\ and\ \bibinfo {author} {\bibfnamefont {A.}~\bibnamefont {Malik}},\
  }\href {\doibase 10.1016/0771-050X(80)90039-X} {\bibfield  {journal}
  {\bibinfo  {journal} {Journal of Computational and Applied Mathematics}\
  }\textbf {\bibinfo {volume} {6}},\ \bibinfo {pages} {295 } (\bibinfo {year}
  {1980})}\BibitemShut {NoStop}%
\bibitem [{\citenamefont {Berntsen}\ \emph {et~al.}(1991)\citenamefont
  {Berntsen}, \citenamefont {Espelid},\ and\ \citenamefont
  {Genz}}]{Berntsen:1991:AAA:210232.210233}%
  \BibitemOpen
  \bibfield  {author} {\bibinfo {author} {\bibfnamefont {J.}~\bibnamefont
  {Berntsen}}, \bibinfo {author} {\bibfnamefont {T.~O.}\ \bibnamefont
  {Espelid}}, \ and\ \bibinfo {author} {\bibfnamefont {A.}~\bibnamefont
  {Genz}},\ }\href {\doibase http://doi.acm.org/10.1145/210232.210233}
  {\bibfield  {journal} {\bibinfo  {journal} {ACM Trans. Math. Softw.}\
  }\textbf {\bibinfo {volume} {17}},\ \bibinfo {pages} {437} (\bibinfo {year}
  {1991})}\BibitemShut {NoStop}%
\bibitem [{Note1()}]{Note1}%
  \BibitemOpen
  \bibinfo {note} {\protect \url
  {http://ab-initio.mit.edu/wiki/index.php/Cubature}}\BibitemShut {NoStop}%
\bibitem [{\citenamefont {Galassi}\ \emph {et~al.}(2003)\citenamefont
  {Galassi}, \citenamefont {Davies}, \citenamefont {Theiler}, \citenamefont
  {Gough}, \citenamefont {Jungman}, \citenamefont {Booth},\ and\ \citenamefont
  {Rossi}}]{GSL}%
  \BibitemOpen
  \bibfield  {author} {\bibinfo {author} {\bibfnamefont {M.}~\bibnamefont
  {Galassi}}, \bibinfo {author} {\bibfnamefont {J.}~\bibnamefont {Davies}},
  \bibinfo {author} {\bibfnamefont {J.}~\bibnamefont {Theiler}}, \bibinfo
  {author} {\bibfnamefont {B.}~\bibnamefont {Gough}}, \bibinfo {author}
  {\bibfnamefont {G.}~\bibnamefont {Jungman}}, \bibinfo {author} {\bibfnamefont
  {M.}~\bibnamefont {Booth}}, \ and\ \bibinfo {author} {\bibfnamefont
  {F.}~\bibnamefont {Rossi}},\ }\href {http://www.worldcat.org/oclc/52055634}
  {\emph {\bibinfo {title} {{Gnu Scientific Library: Reference Manual}}}}\
  (\bibinfo  {publisher} {Network Theory Ltd.},\ \bibinfo {year}
  {2003})\BibitemShut {NoStop}%
\bibitem [{\citenamefont {{Wilde}}(2011)}]{2011arXiv1106.1445W}%
  \BibitemOpen
  \bibfield  {author} {\bibinfo {author} {\bibfnamefont {M.~M.}\ \bibnamefont
  {{Wilde}}},\ }\href@noop {} {\bibfield  {journal} {\bibinfo  {journal} {ArXiv
  e-prints}\ } (\bibinfo {year} {2011})},\ \Eprint
  {http://arxiv.org/abs/1106.1445} {arXiv:1106.1445 [quant-ph]} \BibitemShut
  {NoStop}%
\bibitem [{\citenamefont {{Khoon Ng}}\ and\ \citenamefont
  {{Mandayam}}(2009)}]{2009arXiv0909.0931K}%
  \BibitemOpen
  \bibfield  {author} {\bibinfo {author} {\bibfnamefont {H.}~\bibnamefont
  {{Khoon Ng}}}\ and\ \bibinfo {author} {\bibfnamefont {P.}~\bibnamefont
  {{Mandayam}}},\ }\href@noop {} {\bibfield  {journal} {\bibinfo  {journal}
  {ArXiv e-prints}\ } (\bibinfo {year} {2009})},\ \Eprint
  {http://arxiv.org/abs/0909.0931} {arXiv:0909.0931 [quant-ph]} \BibitemShut
  {NoStop}%
\bibitem [{\citenamefont {{Schumacher}}\ and\ \citenamefont
  {{Westmoreland}}(2001)}]{2001quant.ph.12106S}%
  \BibitemOpen
  \bibfield  {author} {\bibinfo {author} {\bibfnamefont {B.}~\bibnamefont
  {{Schumacher}}}\ and\ \bibinfo {author} {\bibfnamefont {M.~D.}\ \bibnamefont
  {{Westmoreland}}},\ }\href@noop {} {\bibfield  {journal} {\bibinfo  {journal}
  {ArXiv e-prints}\ } (\bibinfo {year} {2001})},\ \Eprint
  {http://arxiv.org/abs/arXiv:quant-ph/0112106} {arXiv:quant-ph/0112106}
  \BibitemShut {NoStop}%
\bibitem [{\citenamefont {{Leung}}\ \emph {et~al.}(1997)\citenamefont
  {{Leung}}, \citenamefont {{Nielsen}}, \citenamefont {{Chuang}},\ and\
  \citenamefont {{Yamamoto}}}]{1997PhRvA..56.2567L}%
  \BibitemOpen
  \bibfield  {author} {\bibinfo {author} {\bibfnamefont {D.~W.}\ \bibnamefont
  {{Leung}}}, \bibinfo {author} {\bibfnamefont {M.~A.}\ \bibnamefont
  {{Nielsen}}}, \bibinfo {author} {\bibfnamefont {I.~L.}\ \bibnamefont
  {{Chuang}}}, \ and\ \bibinfo {author} {\bibfnamefont {Y.}~\bibnamefont
  {{Yamamoto}}},\ }\href {\doibase 10.1103/PhysRevA.56.2567} {\bibfield
  {journal} {\bibinfo  {journal} {\pra}\ }\textbf {\bibinfo {volume} {56}},\
  \bibinfo {pages} {2567} (\bibinfo {year} {1997})},\ \Eprint
  {http://arxiv.org/abs/arXiv:quant-ph/9704002} {arXiv:quant-ph/9704002}
  \BibitemShut {NoStop}%
\bibitem [{\citenamefont {{Taghavi}}\ \emph {et~al.}(2008)\citenamefont
  {{Taghavi}}, \citenamefont {{Kosut}},\ and\ \citenamefont
  {{Lidar}}}]{2008arXiv0810.2524T}%
  \BibitemOpen
  \bibfield  {author} {\bibinfo {author} {\bibfnamefont {S.}~\bibnamefont
  {{Taghavi}}}, \bibinfo {author} {\bibfnamefont {R.~L.}\ \bibnamefont
  {{Kosut}}}, \ and\ \bibinfo {author} {\bibfnamefont {D.~A.}\ \bibnamefont
  {{Lidar}}},\ }\href@noop {} {\bibfield  {journal} {\bibinfo  {journal} {ArXiv
  e-prints}\ } (\bibinfo {year} {2008})},\ \Eprint
  {http://arxiv.org/abs/0810.2524} {arXiv:0810.2524 [quant-ph]} \BibitemShut
  {NoStop}%
\bibitem [{\citenamefont {Smith}\ and\ \citenamefont
  {Raymer}(2006)}]{PhysRevA.74.062104}%
  \BibitemOpen
  \bibfield  {author} {\bibinfo {author} {\bibfnamefont {B.~J.}\ \bibnamefont
  {Smith}}\ and\ \bibinfo {author} {\bibfnamefont {M.~G.}\ \bibnamefont
  {Raymer}},\ }\href {\doibase 10.1103/PhysRevA.74.062104} {\bibfield
  {journal} {\bibinfo  {journal} {Phys. Rev. A}\ }\textbf {\bibinfo {volume}
  {74}},\ \bibinfo {pages} {062104} (\bibinfo {year} {2006})}\BibitemShut
  {NoStop}%
\bibitem [{\citenamefont {{Hamadou Ibrahim}}\ \emph {et~al.}(2012)\citenamefont
  {{Hamadou Ibrahim}}, \citenamefont {{Roux}}, \citenamefont {{Goyal}},
  \citenamefont {{McLaren}}, \citenamefont {{Konrad}},\ and\ \citenamefont
  {{Forbes}}}]{2012arXiv1210.2867H}%
  \BibitemOpen
  \bibfield  {author} {\bibinfo {author} {\bibfnamefont {A.}~\bibnamefont
  {{Hamadou Ibrahim}}}, \bibinfo {author} {\bibfnamefont {F.~S.}\ \bibnamefont
  {{Roux}}}, \bibinfo {author} {\bibfnamefont {S.}~\bibnamefont {{Goyal}}},
  \bibinfo {author} {\bibfnamefont {M.}~\bibnamefont {{McLaren}}}, \bibinfo
  {author} {\bibfnamefont {T.}~\bibnamefont {{Konrad}}}, \ and\ \bibinfo
  {author} {\bibfnamefont {A.}~\bibnamefont {{Forbes}}},\ }\href@noop {}
  {\bibfield  {journal} {\bibinfo  {journal} {ArXiv e-prints}\ } (\bibinfo
  {year} {2012})},\ \Eprint {http://arxiv.org/abs/1210.2867} {arXiv:1210.2867
  [physics.optics]} \BibitemShut {NoStop}%
\bibitem [{\citenamefont {Malik}\ \emph {et~al.}(2012)\citenamefont {Malik},
  \citenamefont {O'Sullivan}, \citenamefont {Rodenburg}, \citenamefont
  {Mirhosseini}, \citenamefont {Leach}, \citenamefont {Lavery}, \citenamefont
  {Padgett},\ and\ \citenamefont {Boyd}}]{Malik:12}%
  \BibitemOpen
  \bibfield  {author} {\bibinfo {author} {\bibfnamefont {M.}~\bibnamefont
  {Malik}}, \bibinfo {author} {\bibfnamefont {M.}~\bibnamefont {O'Sullivan}},
  \bibinfo {author} {\bibfnamefont {B.}~\bibnamefont {Rodenburg}}, \bibinfo
  {author} {\bibfnamefont {M.}~\bibnamefont {Mirhosseini}}, \bibinfo {author}
  {\bibfnamefont {J.}~\bibnamefont {Leach}}, \bibinfo {author} {\bibfnamefont
  {M.~P.~J.}\ \bibnamefont {Lavery}}, \bibinfo {author} {\bibfnamefont {M.~J.}\
  \bibnamefont {Padgett}}, \ and\ \bibinfo {author} {\bibfnamefont {R.~W.}\
  \bibnamefont {Boyd}},\ }\href {\doibase 10.1364/OE.20.013195} {\bibfield
  {journal} {\bibinfo  {journal} {Opt. Express}\ }\textbf {\bibinfo {volume}
  {20}},\ \bibinfo {pages} {13195} (\bibinfo {year} {2012})}\BibitemShut
  {NoStop}%
\bibitem [{\citenamefont {{Rodenburg}}\ \emph {et~al.}(2013)\citenamefont
  {{Rodenburg}}, \citenamefont {{Mirhosseini}}, \citenamefont {{Malik}},
  \citenamefont {{Yanakas}}, \citenamefont {{Maher}}, \citenamefont
  {{Steinhoff}}, \citenamefont {{Tyler}},\ and\ \citenamefont
  {{Boyd}}}]{2013arXiv1301.7454R}%
  \BibitemOpen
  \bibfield  {author} {\bibinfo {author} {\bibfnamefont {B.}~\bibnamefont
  {{Rodenburg}}}, \bibinfo {author} {\bibfnamefont {M.}~\bibnamefont
  {{Mirhosseini}}}, \bibinfo {author} {\bibfnamefont {M.}~\bibnamefont
  {{Malik}}}, \bibinfo {author} {\bibfnamefont {M.}~\bibnamefont {{Yanakas}}},
  \bibinfo {author} {\bibfnamefont {L.}~\bibnamefont {{Maher}}}, \bibinfo
  {author} {\bibfnamefont {N.~K.}\ \bibnamefont {{Steinhoff}}}, \bibinfo
  {author} {\bibfnamefont {G.~A.}\ \bibnamefont {{Tyler}}}, \ and\ \bibinfo
  {author} {\bibfnamefont {R.~W.}\ \bibnamefont {{Boyd}}},\ }\href@noop {}
  {\bibfield  {journal} {\bibinfo  {journal} {ArXiv e-prints}\ } (\bibinfo
  {year} {2013})},\ \Eprint {http://arxiv.org/abs/1301.7454} {arXiv:1301.7454
  [physics.optics]} \BibitemShut {NoStop}%
\end{thebibliography}%

\end{document}